\documentclass[
reprint,
 amsmath,amssymb,
 aps,
]{revtex4-2}

\usepackage{graphicx}% Include figure files
\usepackage{dcolumn}% Align table columns on decimal point
\usepackage{bm}% bold math
\usepackage{upgreek}
\usepackage{amsmath}
\usepackage{mathrsfs}
\usepackage{mathtools}
\usepackage{xcolor}
\usepackage{cases}
\usepackage{ifthen}
\usepackage{soul}
\newcommand{\fpm}{f_{\mathrm{PM}}}

\newcommand{\abs}[1]{\left|#1\right|}

\newcommand{\micron}{\upmu\mathrm{m}}

\newcommand{\deff}{d_{\mathrm{eff}}}
\newcommand{\lcr}{L_{\mathrm{cr}}}
\newcommand{\lcav}{L_{\mathrm{cav}}}
\newcommand{\knet}{k^{''}_{\mathrm{net}}} % beta2 net
\newcommand{\dtil}{\tilde{\delta}}

\newcommand{\trt}{t_\mathrm{rt}}
\newcommand{\dk}{\Delta k}
\newcommand{\dkp}{\Delta k^{\prime}}
\newcommand{\sinc}[1]{\mathrm{sinc}\left( #1 \right)}

\newcommand{\fourier}[1]{\mathcal{F}\left\lbrace #1 \right\rbrace}
\newcommand{\invfourier}[1]{\mathcal{F}^{-1}\left\lbrace #1 \right\rbrace}

\newcommand{\gp}[1]{g_{\mathrm{P}}(#1)}
\newcommand{\Bin}{B_{\mathrm{in}}}

\newcommand{\ppm}{\phi_{\mathrm{PM}}}
\newcommand{\fsr}{\Delta\nu}

\newcommand{\zin}{z_{\mathrm{cr}}^{\mathrm{in}}}
\newcommand{\zout}{z_{\mathrm{cr}}^{\mathrm{out}}}

\usepackage{algorithmic}
\usepackage{algorithm}
\usepackage{array}
\usepackage{verbatim}
\usepackage{mathtools}
\usepackage{soul}
\usepackage{cases}
\usepackage{tikz}
\usepackage{pgfplots,pgfplotstable}
\usepackage{tkz-euclide}
\usepackage{siunitx}
\usepackage[flushleft]{threeparttable}
\usetikzlibrary{babel} %%% <--- Don't forget
\usetikzlibrary{arrows,shapes,backgrounds,petri,calc,fadings}
\usetikzlibrary{automata,positioning}
\usetikzlibrary{decorations.markings,decorations.pathmorphing,decorations.pathreplacing}
\usepgfplotslibrary{groupplots,units,fillbetween}
\usepackage{upgreek}
\usepackage{caption}
\pgfplotsset{every axis/.append style={
                    label style={font=\large},
                    }}
\pgfplotsset{compat=1.15}
\definecolor{mynaranja}{rgb}{0.8471,0.3216,0.0941}
\definecolor{mygreen2}{rgb}{0.0,0.71,0.0}
\definecolor{myrojo2}{rgb}{0.71,0 ,0}

\newboolean{EDITEDCOLOR}
\setboolean{EDITEDCOLOR}{false} % Change to "false" for black version
\newcommand{\edited}[1]{%
    \ifthenelse{\boolean{EDITEDCOLOR}}%
        {\textbf{\textcolor{myrojo2}{#1}}}%
        {\textcolor{black}{#1}}%
}

\begin{document}

\preprint{APS/123-QED}

\title{Quadratic frequency comb based on phase-modulated cw-driven optical parametric oscillator with intracavity dispersion control}

\author{A. D. Sanchez}
\email{alfredo.sanchez@icfo.eu}
\homepage{https://github.com/alfredos84}
\affiliation{ICFO-Institut de Ciencies Fotoniques, Mediterranean Technology Park, 08860 Castelldefels, Barcelona, Spain.}

\author{S. Chaitanya Kumar}

\affiliation{Tata Institute of Fundamental Research Hyderabad, 36/P Gopanpally, Hyderabad 500046, Telangana, India.}

\author{M. Ebrahim-Zadeh}
\affiliation{ICFO-Institut de Ciencies Fotoniques, Mediterranean Technology Park, 08860 Castelldefels, Barcelona, Spain.}
\affiliation{Instituciò Catalana de Recerca i Estudis Avancats (ICREA), Passeig Lluis Companys 23, Barcelona 08010, Spain.}

% \collaboration{CLEO Collaboration}%\noaffiliation

\date{\today}% It is always \today, today,

\begin{abstract}
We report a novel configuration of bulk $\chi^{(2)}$ optical parametric oscillator (OPO) capable of delivering coherent broadband phase-locked spectrum when driven by a continuous-wave (cw) pump laser. By deploying an electro-optic modulator (EOM) internal to a degenerate cw OPO based on MgO:sPPLT or MgO:PPLN and implementing intracavity dispersion control, we show that output spectra extending over 9 nm (119 nm) with well-defined phase coherence can be generated. Pumping the cw OPO at 532~nm (1550~nm) for degenerate operation in the normal (anomalous) dispersion regime at 1064~nm (3100~nm), and using the coupled-wave equations to simulate the OPO in the presence of intracavity dispersion compensation, we demonstrate that this device offers a new alternative for $\chi^{(2)}$-based frequency comb generation. We also show that in the time domain the output of such a device corresponds to femtosecond quadratic solitons in both dispersion regimes. The described concept is generic, paving the way for the realization of a new class of coherent broadband frequency comb sources in different spectral regions based on $\chi^{(2)}$ OPOs pumped by cw lasers.
\end{abstract}

\maketitle

%%%%%%%%%%%%%%%%%%%%%%%%%%%% INTRODUCTION %%%%%%%%%%%%%%%%%%%%%%%%%%%%%%%%%
\section{Introduction}
The advent of optical frequency combs has provided a powerful new tool for a diverse range of applications across multiple disciplines, from frequency metrology to high-field physics, spectroscopy, remote sensing, precision distance measurement, non-invasive disease diagnosis, cancer detection, breath analysis, waveform synthesis, optical communications, astronomy, and several still-evolving research fields~\cite{hansch2006nobel}. The most widely established techniques for the generation of frequency combs are based on the use of mode-locked ultrafast lasers~\cite{jones2000carrier} and Kerr microresonators exploiting $\chi^{(3)}$ nonlinear four-wave mixing~\cite{kippenberg2011microresonator}. In recent years, alternative techniques for frequency comb generation based on $\chi^{(2)}$ nonlinearity in optical parametric oscillators (OPOs) have been explored~\cite{mosca2018modulation}. These include experimental demonstration of passive methods in bulk ultrafast OPOs synchronously-driven by femtosecond pump lasers~\cite{marandi2012coherence,ramaiah2013self} pump lasers and bulk cw OPOs exploiting cascaded $\chi^{(2)}:\chi^{(2)}$ nonlinearities~\cite{ulvila2014high}, as well as theoretical study of bulk microring cw OPOs in the presence of $\chi^{(2)}$ and $\chi^{(3)}$ nonlinearity~\cite{villois2019frequency}. On the other hand, the exploitation of active techniques for frequency comb generation has been explored using electro-optic phase modulation in a bulk degenerate cw-driven OPO, where broadband spectral emission has was demonstrated~\cite{diddams1999broadband}. In later work, we investigated the characteristics of such a device in the time domain, where pulses of picosecond temporal duration were generated~\cite{devi2013directly}. However, in all of the above experimental or theoretical studies, passive or active, the role of dispersion in cw-driven OPOs for frequency comb generation was not considered and the influence of this parameter on the spectral and temporal characteristics of the output was not explored. Although the concept of group velocity dispersion compensation, using for example intracavity prism pair or chirped mirrors, is well established in synchronously-pumped femtosecond and picosecond bulk OPOs and ultrafast lasers, or in dispersion-engineered microcavities exploiting $\chi^{(3)}$-based Kerr nonlinearity~\cite{yang2016broadband}, group velocity dispersion control is not a common practice in bulk cw OPOs. To the best of our knowledge, this technique has not been previously investigated and studied in the context of bulk cw-driven OPOs.

In this paper, we present a theoretical framework which describes the comb properties of a phase-modulated cw-driven bulk degenerate OPO with dispersion control. In a three-wave interaction within a quadratic medium, owing to the phase-matching properties of the nonlinear material, the driving pump frequency is significantly far from the generated signal/idler frequencies, making group velocity mismatch (GVM) and group velocity dispersion (GVD) management in these systems not only critical, but also indispensable for comb generation. We show that the implementation of dispersion compensation in such a device is pivotal to achieving coherence properties as a frequency comb with a broadband phase-locked spectrum, with the comb exhibiting similar characteristics in both normal and anomalous dispersion regimes. We also demonstrate that in the time domain the output of such a device corresponds to near-transform-limited solitonic pulses, again with similar characteristics in both dispersion regimes. Our model is based on the well-established coupled-wave equations (CWEs) with proper boundary conditions, which in turn was numerically simulated using the split-step Fourier method (SSFM) scripted in CUDA to leverage the speed up provided by GPU cards~\cite{sanchez2024cuda}. In order to demonstrate our device can operate in both normal and anomalous dispersion regimes, we simulate pumping the OPO at 532 nm (1550 nm) for degenerate operation in the normal (anomalous) dispersion regimes. The generic nature of this active phase modulation scheme is demonstrated using  MgO-doped periodically-poled LiTaO$_3$ and MgO-doped periodically-poled LiNbO$_3$ as the gain media in the normal and anomalous dispersion regions respectively, using a traveling wave ring cavity configuration together with an intracavity EOM. We focus our attention on a degenerate doubly-resonant OPO, since this configuration delivers the broadest output spectrum which is advantageous for numerous applications such as broadband spectroscopy and comb generation as well as enabling the attainment of shortest optical pulses. The paper is organized as follows. In Section~\ref{sec:theory} we present the theoretical model along with details on the numerical implementation. We then present detailed analysis on the operation of our device in temporal and spectral domains in Sections~\ref{sec:results}. Finally, we present some remarkable conclusions.

%%%%%%%%%%%%%%%%%%%%%%%%%% THEORETICAL FRAMEWORK %%%%%%%%%%%%%%%%%%%%%%%%%%%%%
\section{Theoretical framework and numerical implementation}
\label{sec:theory}

\begin{figure*}[ht]
    \centering
    \includegraphics[width=1.0\textwidth]{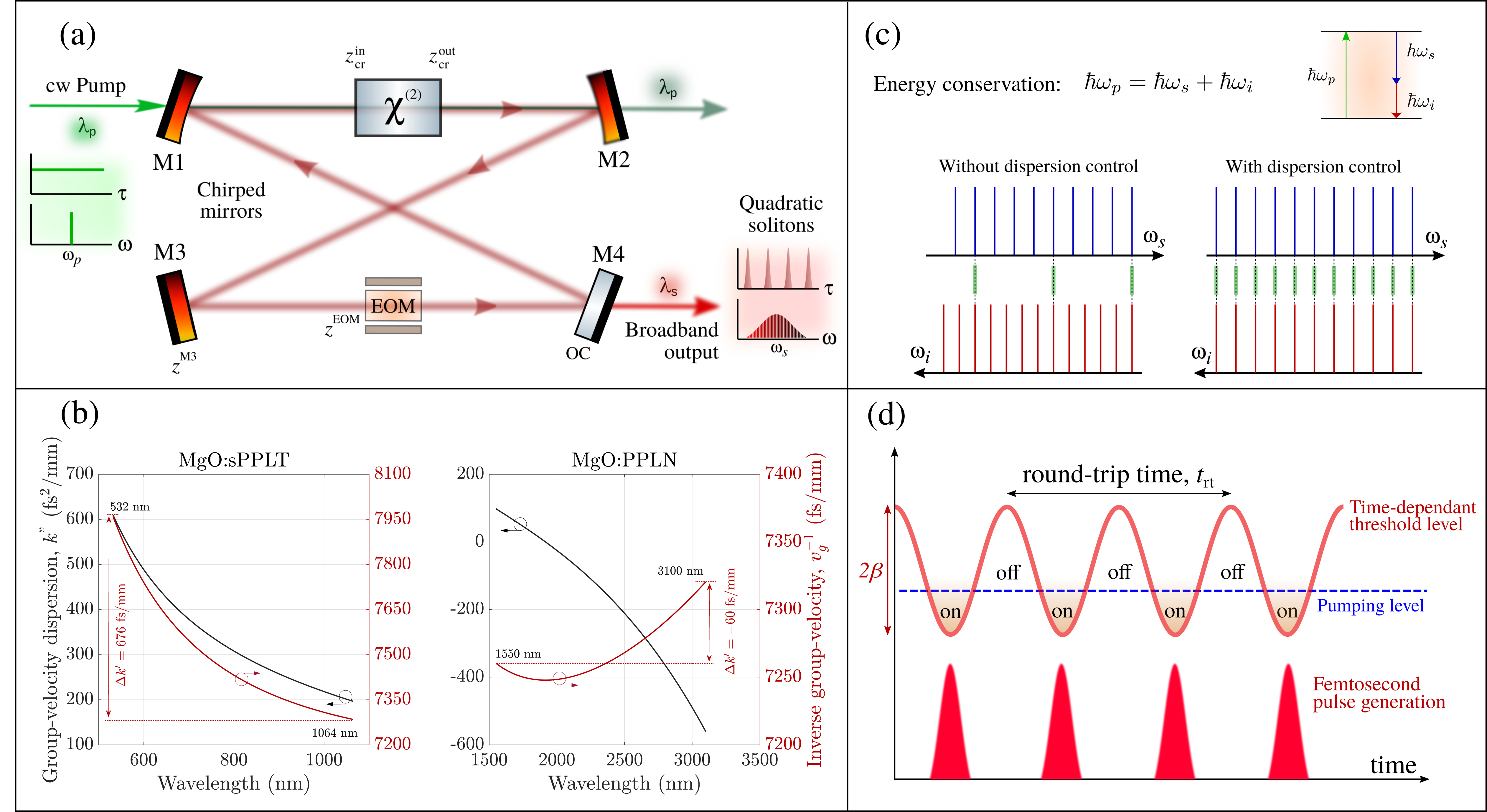}
    \caption{Generic design of the phase-modulated degenerate OPO used in this work. (a) A cw input laser field pumps the OPO at 532~nm (1550~nm) to generate the resonant degenerate signal field at 1064~nm (3100~nm) in the normal (anomalous) dispersion regime by parametric-down conversion in the $\chi^{(2)}$ medium. The intracavity EOM together with dispersion control (DC) of the cavity enables phase-locking for the generation of broadband comb in frequency domain and femtosecond soliton pulses in time domain. (b) Dispersion properties of the selected nonlinear crystals. (c) Energy diagram of the process and the cavity modes without (left) and with (right) dispersion compensation, showing the pair of modes that satisfy the energy conservation (pairs indicated with a green vertical bar in the middle of axes). (d) Schematic of the phase modulation process. The EOM modifies the OPO threshold condition during the round-trip time (red sinusoidal curve). For a fixed pumping level (blue dashed line), the threshold function can switch on/off the OPO in a short time window of the full round trip time. In combination with dispersion control, the system can deliver femtosecond pulses when the OPO is switched on.}
    \label{fig:Fig1}
\end{figure*}

The schematic of the phase-modulated degenerate cw OPO used in our theoretical model is shown in Fig.~\ref{fig:Fig1}(a). A cw single-frequency laser operating at $\lambda_p=532$~nm or 1550~nm is used as the pump source. The OPO is configured in a travelling-wave cavity comprising two concave mirrors (M1, M2) and two plane mirrors (M3, M4). All mirrors are highly transparent for the pump at $\lambda_p$ and highly reflecting for the degenerate signal at $\lambda_s$, except M4, which serves as an output coupler ($R= 98\%$) at the degenerate wavelength. The nonlinear gain medium for the OPO is a periodically-poled crystal with a grating period, $\Lambda$, for type-0 ($e \rightarrow ee$) quasi-phase-matched (QPM) interaction, placed internal to the optical cavity between M1 and M2. The pump beam is focused to a waist radius of $w_{0p}=55$~$\micron$. The round-trip cavity length is 60~mm, corresponding to a free-spectral-range (FSR) of $\fsr\approx 3.9$~GHz. Figure.~\ref{fig:Fig1}(b) shows the dispersion properties of two prominent nonlinear crystals used in our model. The left axis (black) corresponds to group-velocity dispersion, showing that the signal is generated in the normal (MgO:sPPLT, $\lambda_p/\lambda_s=532/1064$~nm) and anomalous (MgO:PPLN, $\lambda_p/\lambda_s=1550/3100$~nm) dispersion regimes, while the right axis (red) indicates the corresponding group velocity as a function of the wavelength~\cite{bruner2003temperature,gayer2008temperature}. The group-velocity mismatch (GVM) is 676~fs/mm and -60~fs/mm for MgO:sPPLT and MgO:PPLN, respectively.

The net intracavity GVD can be controlled by using chirped cavity mirrors (M1, M2, and M3) with suitable group-delay dispersion (GDD) profile, intracavity dispersive elements or wedges. Figure~\ref{fig:Fig1}(c) depicts the effect of the dispersion compensation in a doubly-resonant OPO cavity. Without dispersion compensation (left panel), only a few modes satisfy the energy conservation condition due to the Vernier effect between the signal and idler FSRs, $\fsr_s$ and $\fsr_i$. Hence, after compensating for the net intracavity dispersion, the Vernier effect will be removed, enabling spectral broadening (right panel).

Similar to mode-locked lasers, the EOM induces a coupling between adjacent modes, locking their phases, and shortening the pulses within the gain medium bandwidth~\cite{kuizenga1970fm}. The difference with the mode-locked laser is that the gain in an OPO is instantaneous, a consequence of the nonlinear polarization induced in the nonlinear crystal. In our scheme, an EOM placed between mirrors M3 and M4 provides a sinusoidal phase modulation function $\ppm(\tau)=\beta\cos{(2\pi\fpm \tau)}$ to the intracavity degenerate field for active phase modulation of the degenerate OPO. Here, $\beta$ is the modulation depth and $\fpm$ the modulation frequency, which is precisely set to synchronize with the cavity FSR (Fig.~\ref{fig:Fig1}(c)). This phase modulation has the effect of modifying the OPO threshold condition during the round-trip time~\cite{sanchez2024mean}, allowing the OPO to oscillate only in a short time window. This is sketched in Fig.~\ref{fig:Fig1}(d) where the EOM-dependant threshold function (red curve) turns on/off the OPO depending upon the pumping level (blue dashed horizontal line). As we will show later (see Sec.~\ref{sec:tempdomain}), in the time domain, this modulation of pumping level above and below the threshold level, together with net cavity dispersion compensation, leads to the generation of femtosecond pulses.

We model the evolution of the degenerate signal electric field using the well-known coupled-wave equations for three-wave interactions under the plane-wave approximation~\cite{AlfredoJSTQE2022}. For the chosen set of parameters, the resulting focusing parameter, $\xi<1$, justifies this approximation. For degenerate parametric-down conversion process, with indistinguishable signal and idler, the coupled-wave equations are
\begin{widetext}
\begin{subequations}
\begin{eqnarray}
    \frac{\partial A_{p}^{(m)}}{\partial z} &= -\left( \frac{\alpha_{cp}}{2}+\Delta k^{\prime}\frac{\partial}{\partial \tau}+i\frac{k^{''}_p}{2}\frac{\partial^2}{\partial \tau^2} \right) A_{p}^{(m)} + i\kappa_p A_{s}^{(m)} A_{s}^{(m)}e^{-i\Delta k z} \label{eq:CEp}\\
    \frac{\partial A_{s}^{(m)}}{\partial z} &= -\left( \frac{\alpha_{cs}}{2} +i\frac{k^{''}_s}{2}\frac{\partial^2}{\partial \tau^2} \right) A_{s}^{(m)} + i\kappa_s A_{p}^{(m)} A_{s}^{*(m)}e^{+i\Delta k z} \label{eq:CEs}
\end{eqnarray}
\end{subequations}
\end{widetext}
where $A_{s}^{(m)}(z,\tau)$ and $A_{p}^{(m)}(z,\tau)$ are the corresponding pump and degenerate (signal) fields, respectively, for a single pass over the $m^{\mathrm{th}}$ round-trip, $z$ is the spatial propagation coordinate, which inside the crystal is $z\in\left[\zin, \zout\right]$. 
The temporal variable, $\tau\in\left[-\trt/2,\trt/2\right]$, describes the temporal window during the round-trip time, $\trt$, in a co-moving frame with a signal group velocity, $v_{gs}=(\left.\partial k/\partial \omega\right|_{\omega_s})^{-1}$, $\alpha_{cx}$ ($x=p,~s$) is the crystal linear absorption at pump and signal, $\Delta k^{\prime} = (\left.\partial k/\partial \omega\right|_{\omega_p})^{-1}-(\left.\partial k/\partial \omega\right|_{\omega_s})^{-1}$ is the group-velocity mismatch (GVM), and $k^{''}_x=\left.\partial^2 k/\partial \omega^2\right|_{\omega_x}$ is the group-velocity dispersion (GVD). $\dk = 2k(\omega_s)-k(\omega_p)-2\pi/\Lambda$ represents the phase-mismatch factor, where $\omega_x$ is the angular frequency of the propagating fields, and $\Lambda$ is the QPM grating period. The nonlinear coupling constant is  $\kappa_x=2\pi\deff/n_x\lambda_x$, where $\deff$~and~$n_x$ are the effective nonlinear coefficient and refractive index of the crystal evaluated at the respective wavelengths using the relevant Sellmeier equations~\cite{bruner2003temperature,gayer2008temperature}. 

Before expressing how the electric field of the signal is updated after one round-trip, it is convenient to describe how the intracavity elements modify the resonant field:
\begin{enumerate}
    \item \textbf{Dispersion compensation} consists of adding a frequency-dependent phase in order to cancel the same effect arising from the nonlinear crystal. This is done in the frequency domain: $\tilde{A}_m(z^{\mathrm{M3}},\Omega)\rightarrow\tilde{A}_m(z^{\mathrm{M3}},\Omega)e^{i\frac{\gamma}{2}\Omega^2}$. 
    The parameter, $\gamma$, corresponds to the  GDD compensation factor,
    \begin{equation}
        \gamma = -\epsilon \lcr k^{''}_s,    
    \end{equation}
    where the unitless \textit{compensation index}, $\epsilon\in\left[0,1\right]$, accounts for the crystal GVD compensation. Thus, the net intracavity GVD is $\knet(\epsilon) = \left( 1-\epsilon \right) k^{''}_s$. Here we have considered GVD alone, but higher-order dispersion can also be readily included in the model if needed. 
    \item \textbf{Electro-optic modulator} inclusion implies adding a phase in the time domain. This step is performed in the time domain by firstly taking the inverse Fourier transform of the dispersion-compensated signal field: $A_m(z^{\mathrm{EOM}},\tau)\rightarrow A_m(z^{\mathrm{EOM}},\tau)e^{i\dtil(\tau)}$. It should be noted that free-space propagation does not affect the resonant field. Here, $\dtil$ is the net time-dependant cavity detuning,
    \begin{equation}
      \dtil(\tau)=% 
      \begin{cases}
        \delta+\beta\cos(2\pi\fpm\tau) &\text{if EOM is 'on'} \\
        \delta &\text{if EOM is 'off'}
      \end{cases}\label{eq:Detuning}
    \end{equation}
    The EOM imposes a time-dependant cavity detuning on the resonant signal frequency, $\omega_0$, and $\delta=(\omega_0-\omega_{\mathrm{cav}})\trt$ is the static detuning with respect to the cavity mode, $\omega_{\mathrm{cav}}$. For the sake of simplicity, in what follows, we will consider $\delta=0$.
    \item \textbf{Cavity losses}\\
    The final step is to multiply the resonant electric field by $\sqrt{R}$ in the time domain, before starting the next round-trip. This accounts for the relevant cavity losses for the propagated, GVD-compensated, and phase-modulated resonant field.
\end{enumerate}

Therefore, the boundary conditions for resonant signal field for each round-trip, including the EOM and dispersion compensation, can be written as
\begin{equation}
    A^{(m+1)}_s(0,\tau) = \sqrt{R}e^{i\dtil(\tau)}\invfourier{\tilde{A}^{(m)}_s(\zout,\Omega)e^{i\frac{\gamma}{2}\Omega^2}} \label{eq:BCs}
\end{equation}
where $\tilde{A}^{(m)}(z,\Omega)=\fourier{A^{(m)}(z,\tau)}$ is the Fourier transform defined as $\fourier{\cdot}=\int_{-\infty}^{+\infty}\cdot e^{+i\Omega\tau}d\tau$ and $\Omega$ is the relative angular frequency with respect to the carrier, $R$ is the reflectivity at the signal wavelength, and $z=0$ corresponds to the cavity input position at M1. 

%%%%%%%%%%%%%%%%%%%%% RESULTS %%%%%%%%%%%%%%%%%%%%%%%%%%%%
\section{Results}
\label{sec:results}

In order to study the evolution of the signal field in our phase-modulated degenerate OPO, the set of Eqs.~\ref{eq:CEp},~\ref{eq:CEs}~and~\ref{eq:BCs}, which are usually referred to as an infinite-dimensional map with respect to $\tau$~\cite{buryak2002optical}, should be solved numerically to separately analyze the corresponding solutions in the spectral and temporal domain. A detailed analysis of an OPO with an intracavity EOM, but without dispersion control, can be found in Ref.~\cite{AlfredoJSTQE2022}. This section is divided into two subsections dealing separately with the spectrum and temporal domain. We first analyze the operation of the OPO with EOM+GDD compensation by setting the frequency of the EOM identically to the free-spectral range, $\fpm=\fsr$. This results in the formation of two overlapped coherent and broadband spectra due to the properties of the threshold. The formed spectra exhibit a linear spectral phase relationship between the subsequent spectral lines. As a consequence, in the time domain a double-pulse formation is observed. Second, we fine-tune the $\fpm$ to eliminate one of the generated pulses. This leads to the formation of a coherent phase-locked spectrum with comb properties, whilst in the time domain it results in the generation of a single femtosecond quadratic soliton.

We use the Split-Step Fourier Method (SSFM) to solve this infinite-dimensional map, where the crystal is divided into thin slices and the linear part is solved in the frequency domain, while the nonlinear part is solved in the time domain using a standard fourth-order Runge-Kutta method. Our implementation of SSFM is a highly-parallelizable algorithm since it involves successive Fourier transforms and vector operations during the simulation that are efficiently solved on GPU. For this propose we use our flexible CUDA-based code that models OPOs under different configurations~\cite{sanchez2024cuda, Nvidia}. All the simulations presented in this work are performed over 10000 round-trips to ensure steady-state operation of the OPO, as well as with the net cavity dispersion fully compensated. The parameters used in our simulations are detailed in Table~\ref{tab:Table1}.
\begin{table}%[htbp]
\centering
\footnotesize
\begin{threeparttable}
\caption{\bf Parameters used for our simulations. Sellemeier equations~\cite{bruner2003temperature,gayer2008temperature}.}
\vspace{5mm}
\begin{tabular}{l|r|l}
\hline
\textrm{\textbf{Parameter}} & \textrm{\textbf{Values}$^{(a)}$} & \textrm{\textbf{Units}}\\
\hline
Pump wavelength, $\lambda_p$       & 532/1550     & nm                      \\
Degenerate wavelength, $\lambda_s$ & 1064/3100    & nm                      \\
Pump beam waist, $w_{0p}$          & 55      & $\upmu$m                \\
Cavity net reflectivity, $R$       & 0.98       & -                      \\
Cavity length, $\lcav$             & 60      & mm                      \\
Free spectral range, FSR           & 3.89/3.93    & GHz                     \\
Round trip time, $\trt$            & 256.87/254.60  & ps                      \\ 
\hline
$n_p$                              & 2.20/2.13    & -                       \\
$n_s$                              & 2.13/2.09    & -                       \\
$\dk$                              & $\approx 0$ & $\upmu$m$^{-1}$\\
$\dk^{\prime}$                     & 676/-60     & fs mm$^{-1}$                    \\
$k^{''}_p$                         & 610/98 & fs$^2$mm$^{-1}$  \\
$k^{''}_s$                         & 197/-561 & fs$^2$mm$^{-1}$  \\
$\deff$                            & 7/16   & pm V$^{-1}$                    \\
Crystal length, $\lcr$             & 15      & mm                      \\
Grating period, $\Lambda$          & 7.97/35.01    & $\upmu$m                \\
Crystal temperature                & 57.18/32.88     & $^{\circ}$C                   \\ 
\hline
Time step ($\Delta \tau$)          & 16    & fs                      \\
Step size, $dz$                    & 150     & $\upmu$m                \\
Round trips per simulation         & $10000$   & -                       \\ 
Execution time$^{(b)}$             & $\approx1000$    & round trips min$^{-1}$       \\ 
\hline
\end{tabular}
\begin{tablenotes}
  \item \scriptsize $^{(a)}$ The quantities with two values correspond to the MgO:sPPLT (left) and MgO:PPLN (right).
  \item \scriptsize $^{(b)}$ We used for all simulations a GPU card NVIDIA, model GeForce GTX 1650. However, the execution time varies depending on the GPU used.
\end{tablenotes}
    \label{tab:Table1}
\end{threeparttable}
\end{table}

%%%%%%%%%%%%%%%%%%%%% SUBSECTION SPECTRAL ANALISYS %%%%%%%%%%%%%%%%%%%%%%%%%%%%
\subsection{Frequency Domain: Comb formation in phase-modulated OPO}
\label{sec:freqdomain}

\begin{figure} %[htbp]
    \centering
    \includegraphics[width=\linewidth]{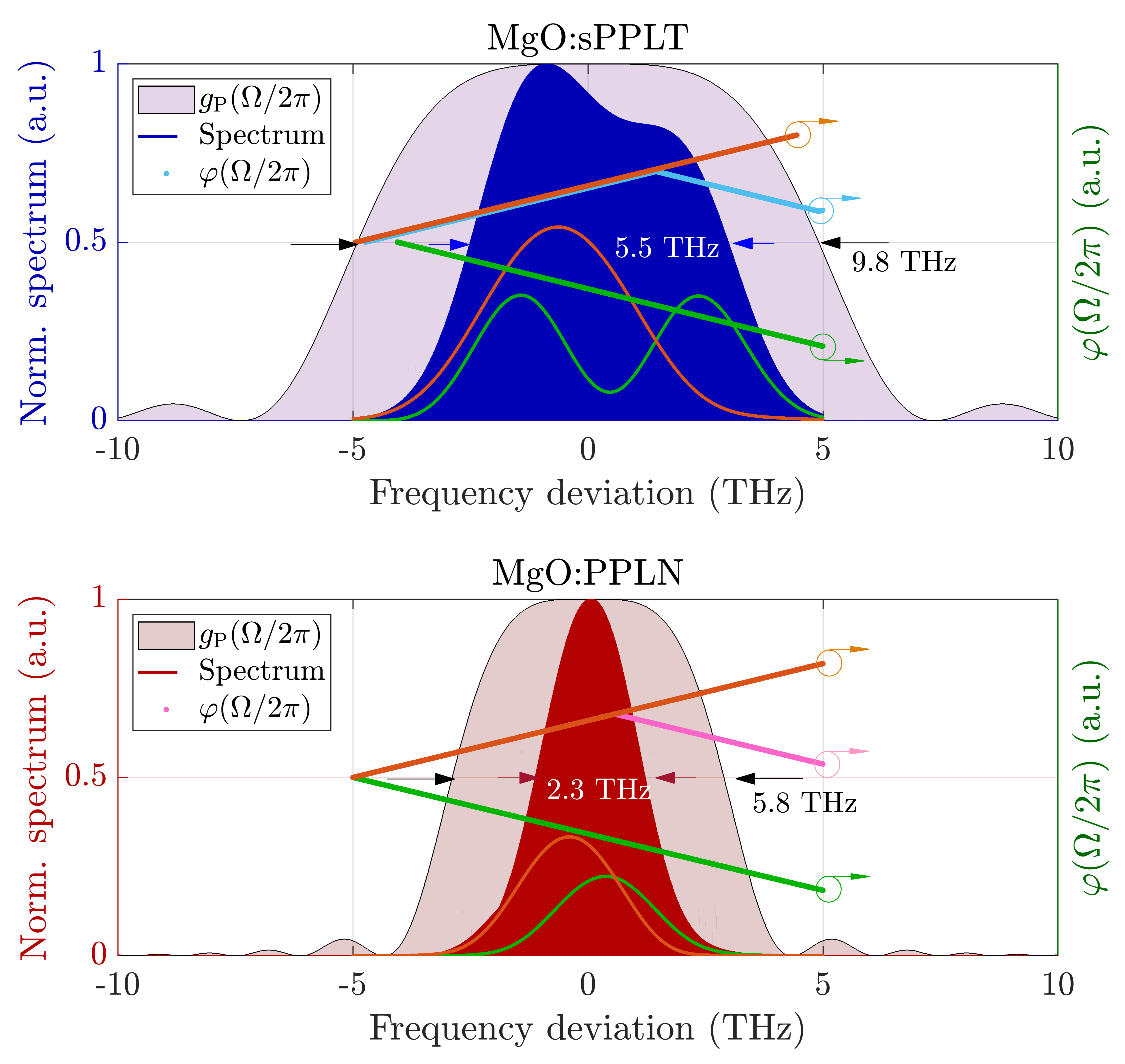}
    \caption{Intracavity two-pulses spectrum formation and spectral phase for $\fpm=\fsr$. The decomposition of the full spectrum into two single-pulses and the corresponding spectral phase, $\varphi$, is also shown.}
    \label{fig:freq2pulses}
\end{figure}

In order to ensure efficient coupling between the neighbouring modes in a phase-modulated OPO, the synchronization between the modulation frequency, $\fpm$, and the FSR of the OPO cavity is critical~\cite{devi2013directly}. The detuning between $\fpm$ and FSR results in an increased threshold for the OPO. Moreover, the phase sensitive nature of the parametric process leads to a dynamic threshold behaviour due to active intracavity phase modulation. Following the calculations developed in Ref.~\cite{sanchez2024mean}, the threshold condition is satisfied if
\begin{equation}\label{eq:N_EOM}
    N \geq 1+\left(\frac{\beta\cos\left(2\pi\fpm\tau\right)}{\alpha_s}\right)^2,
\end{equation}
where $N=I_p/I_{\mathrm{th}}$ represents the pumping level, with $I_p$ and $I_{\mathrm{th}}$ the input pump intensity and the intensity threshold, respectively, and $\alpha_s=\left(1-R+\lcr\alpha_{cs}\right)/2$ accounts for the total cavity loss at the signal wavelength~\cite{mosca2018modulation}. The right-hand side of the inequality in~Eq.~\ref{eq:N_EOM} represents the time-dependant threshold function. For a fixed $N$ value, the threshold condition is satisfied around two windows centered at $\tau=\pm \trt/4$, for $\fpm=\fsr=\trt^{-1}$. Under this condition, there will be two time slots over the full round-trip in which the OPO switches on, allowing the pump-to-signal energy transfer. Figure~\ref{fig:freq2pulses} shows the results of the simulations for both crystals: MgO:sPPLT and MgO:PPLN in the top and bottom panels, respectively. Panels depict the full spectrum (dark shaded curves) along with the single-pulse spectra (solid curves inside the full spectrum), and the parametric gain (light shaded areas) defined as~\cite{ebrahimzadeh2001optical}
\begin{equation}
    \gp{\Omega} = \left[\kappa_s\lcr\abs{\Bin} \sinc{\frac{\dk(\Omega)\lcr}{2}}\right]^2,
\end{equation}
where $\Bin$ is the input electric field at the pump wavelength. In the right vertical axes the corresponding spectral phases, $\varphi(\Omega)$, are also plotted.
The obtained spectral bandwidth is 5.5~THz (2.3~THz) for normal (anomalous) dispersion regime for the MgO:sPPLT (MgO:PPLN) nonlinear crystal. As expected, the spectral phase exhibits a linear variation across the signal spectrum in both crystals. This can be understood from the definition of GDD, that is related to the spectral phase according to $\varphi_2(\Omega)=\partial^2\varphi/\partial\Omega^2$. Since the spectral phase is linear, its second-order derivative is identically zero, being consistent with full intracavity dispersion compensation. Our results show that the phase difference between two adjacent modes, $\Delta\varphi_{\mu}=\varphi_{\mu+1}-\varphi_{\mu}$, where $\mu$ accounts for the signal mode number, takes the values $\Delta\varphi_{\mu}\approx \pm \pi/2$. 

\begin{figure*}[htbp]
    \centering
    \includegraphics[width=0.8\linewidth]{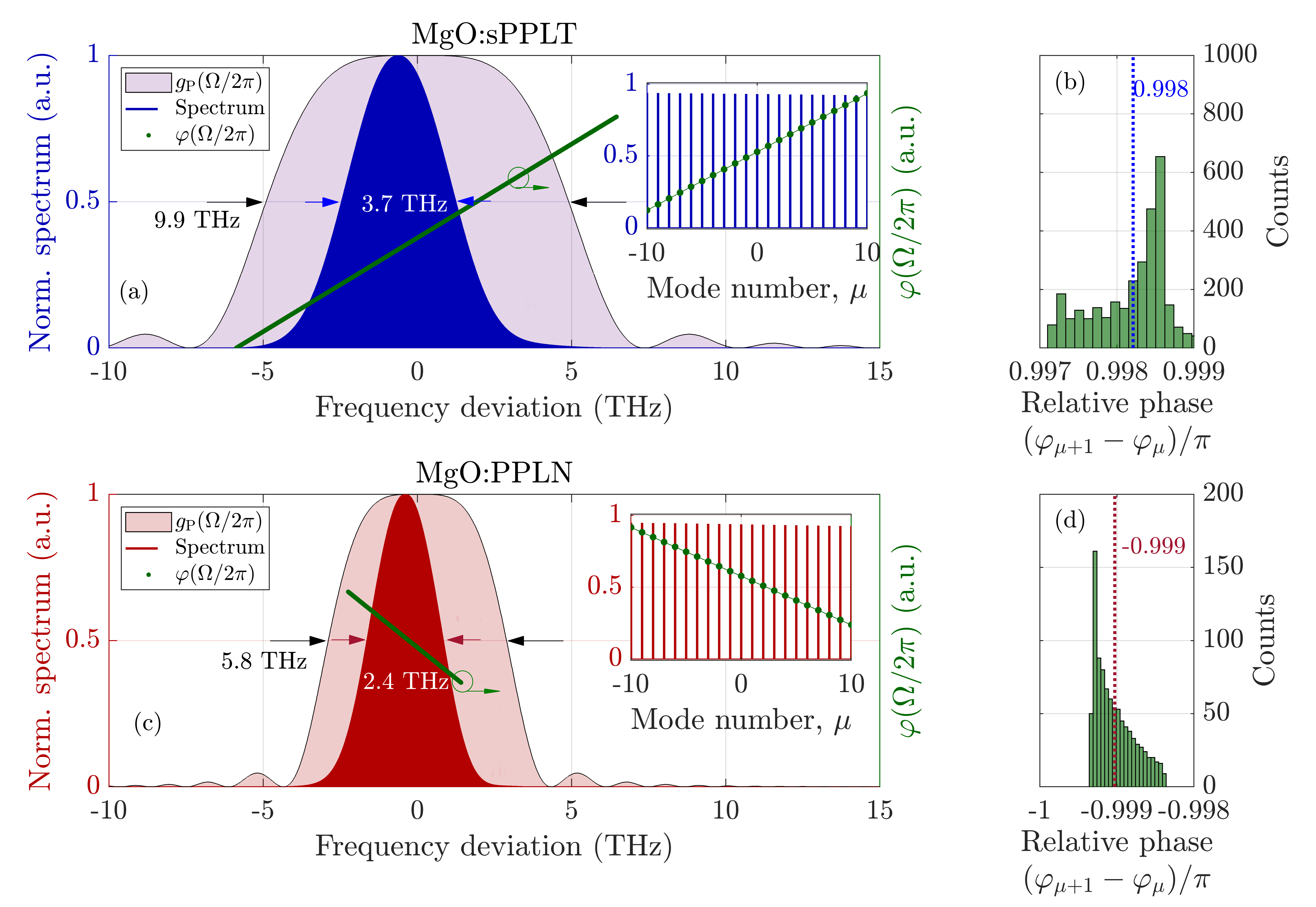}
    \caption{Intracavity single-pulse spectrum formation and spectral phase for $\delta f= \delta f^{\mathrm{LB}}$.}
    \label{fig:freq1pulse}
\end{figure*}

We are interested in finding the conditions in which the double spectrum in Fig.~\ref{fig:freq2pulses} is reduced into a single spectrum. For this propose, we slightly detune the phase modulation frequency to modify the threshold condition in the edges of the round-trip such that a single spectrum is sustained inside the cavity. Let us turn our attention to Eq.~\ref{eq:Detuning} and replace the function $\cos()$ by $\sin()$, for the sake of simplicity. This is completely valid because the effect is to modify the time in which the threshold condition is satisfied. By means of this modification, the threshold condition is verified around $\tau=0$ and $\tau=\pm \trt/2$ (edges of the round-trip). This fine tuning can be obtained from Eq.~\ref{eq:N_EOM} as~\cite{sanchez2024mean}
\begin{equation}
    N \leq 1 + \left(\frac{ \beta\sin\left(\pm \pi\trt\fpm\right) }{\alpha_s}\right)^2,
\end{equation}
and solving for $\delta f = \fsr - \fpm$, we obtain
\begin{widetext}
\begin{equation}
    \frac{\alpha_s}{\beta} \sqrt{N-1} \leq \abs{\sin\left(\pm \pi\trt(\fsr-\delta f)\right)}=\abs{\sin\left(\pm \pi \frac{\delta f}{\fsr}\right)}\approx \abs{\pi \frac{\delta f}{\fsr}}
\end{equation}
\end{widetext}
where we used the property $\sin(a-b)=\sin(a)\cos(b)-\cos(a)\sin(b)$, and the assumption that $\delta f \ll \fsr$. Hence we find that
\begin{equation}\label{eq:dflowerbound}
    \delta f \gtrapprox \frac{\alpha_s\sqrt{N-1}}{\pi\beta}\fsr.
\end{equation}
An estimated lower bound is given by equating Equation~\ref{eq:dflowerbound}
\begin{equation}\label{eq:dflowerbound_estim}
    \delta f^{\mathrm{LB}} = \frac{\alpha_s\sqrt{N-1}}{\pi\beta}\fsr,
\end{equation}
with the factor $\alpha_s\sqrt{N-1}/\pi\beta\sim 10^{-3}$, for $\alpha_s=0.015$, $N = 4$, and $\beta= 0.8\pi$, relevant for our case, suggesting that $\delta f \approx 50$~MHz for $\fsr\approx3.9$~GHz. Figure~\ref{fig:freq1pulse} shows the spectrum for both normal (panel (a)) and anomalous (panel (c)) regimes. The obtained spectral bandwidth is 3.7~THz (2.4~THz) for normal (anomalous) dispersion regime. The insets show the comb formation with a line separation corresponding to the cavity $\fsr$, where the spectral phase is also indicated following a linear tendency, but adopting a value $\Delta\varphi_{\mu}\approx \pm \pi$, as shown in the histograms in panels (b, d). It should be noted that for higher values of $\dkp\lcr$ it is possible to obtain solitonic solutions but with a more complex pulse shape such as a multipeak structure, as reported in Ref.~\cite{nie2020quadratic2}.

%%%%%%%%%%%%%%%%%%%%% SUBSECTION SINGLE PULSE %%%%%%%%%%%%%%%%%%%%%%%%%%%%
\subsection{Time domain: Quadratic soliton generation in phase-modulated OPO}
\label{sec:tempdomain}

\begin{figure*}[htbp]
    \centering
    \includegraphics[width=0.8\linewidth]{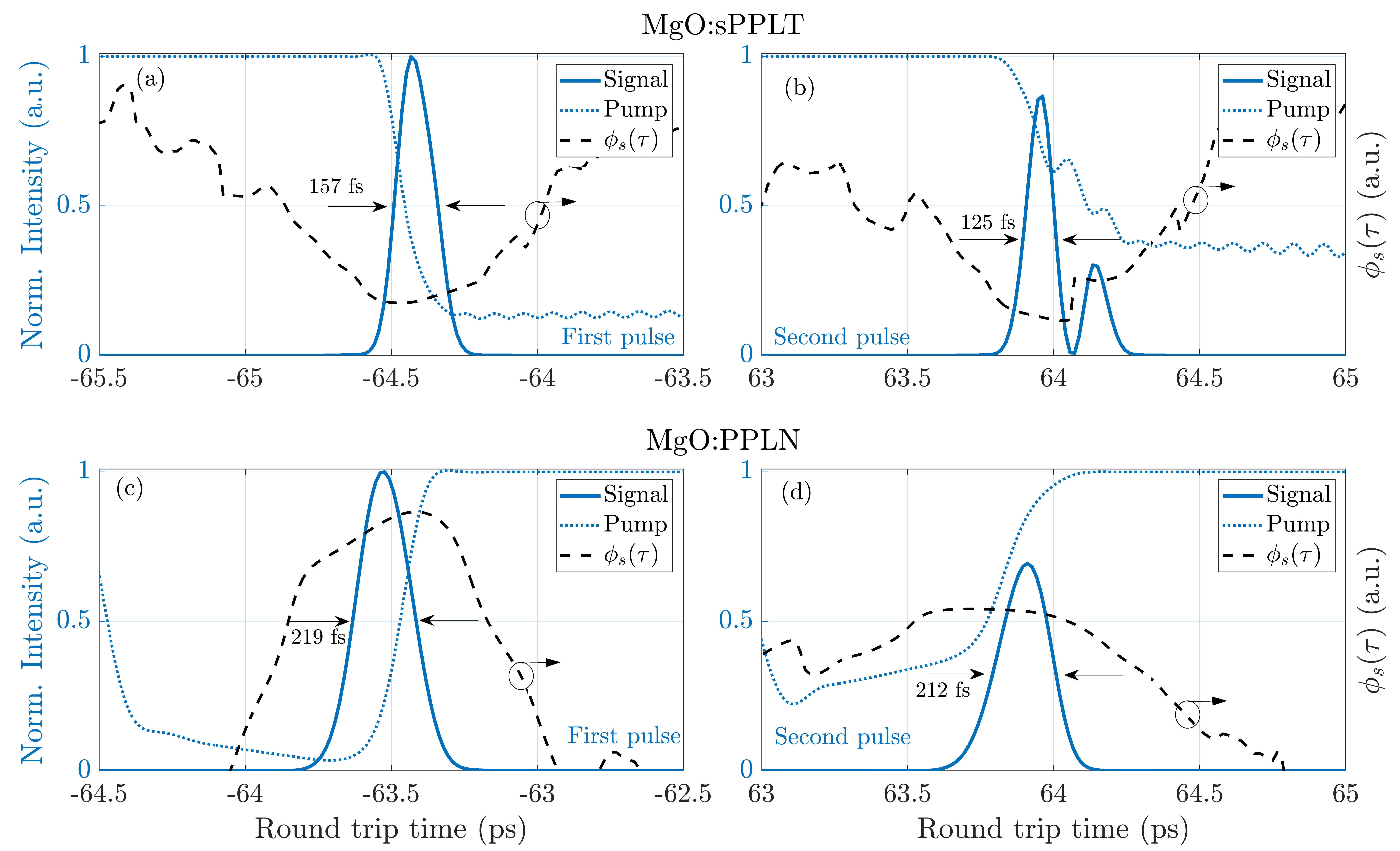}
    \caption{Intracavity pulses formation for $\fpm=\fsr$. The pulses are formed in the two short-slices of the full round-trip time ($\sim 250$~ps) in which the threshold condition is satisfied.}
    \label{fig:time2}
\end{figure*}

Now let us turn our attention to the temporal domain. Figure~\ref{fig:time2} shows the simulation results for MgO:sPPLT (top row) and MgO:PPLN (bottom row). The round-trip time for our simulations is $\sim 250$~ps (see Table~\ref{tab:Table1}). Each panel shows in detail the signal pulse shape and duration (blue solid curve), along with the pump intensity (blue dotted curve) and temporal signal phase (black dashed curve), $\phi_s(\tau)$. For the case of MgO:sPPLT, the pulses durations are 157 and 125~fs at 1064~nm, while for MgO:PPLN the pulses durations are 219 and 212~fs.

There are two interesting features of the obtained pulses. The first is that the pulses are not symmetric, or in the case of panel (b) the obtained pulse has a dual-peak structure. This asymmetry is due to the conversion process and GVM effects. The effect is enhanced in the case of MgO:sPPLT (panel (b)), where $\dkp$ is one order of magnitude higher than in the case of MgO:PPLN. The second is the fact the two intracavity pulses differ from each other because the chirp induced by the EOM has opposite sign at $\tau=-\trt/4$ and $\tau=+\trt/4$~\cite{sanchez2024mean}
\begin{equation}
    \left. \frac{d\phi_s}{d\tau} \right|_{\tau=\pm \trt/4} = \pm \frac{\pi\beta\fpm}{\alpha_s\sqrt{N}}.
\end{equation} 
Based on the values used in our simulations and the resulting pulse duration in Fig.~\ref{fig:time2}, the estimated change in signal temporal phase, $\Delta\phi_s$, in presence of phase-modulation is $\approx 0.3$~rad, being a marginal value taking into account the full net cavity dispersion compensation. This shows the our system can deliver nearly transform-limited pulses.

\begin{figure*}[htbp]
    \centering
    \includegraphics[width=\linewidth]{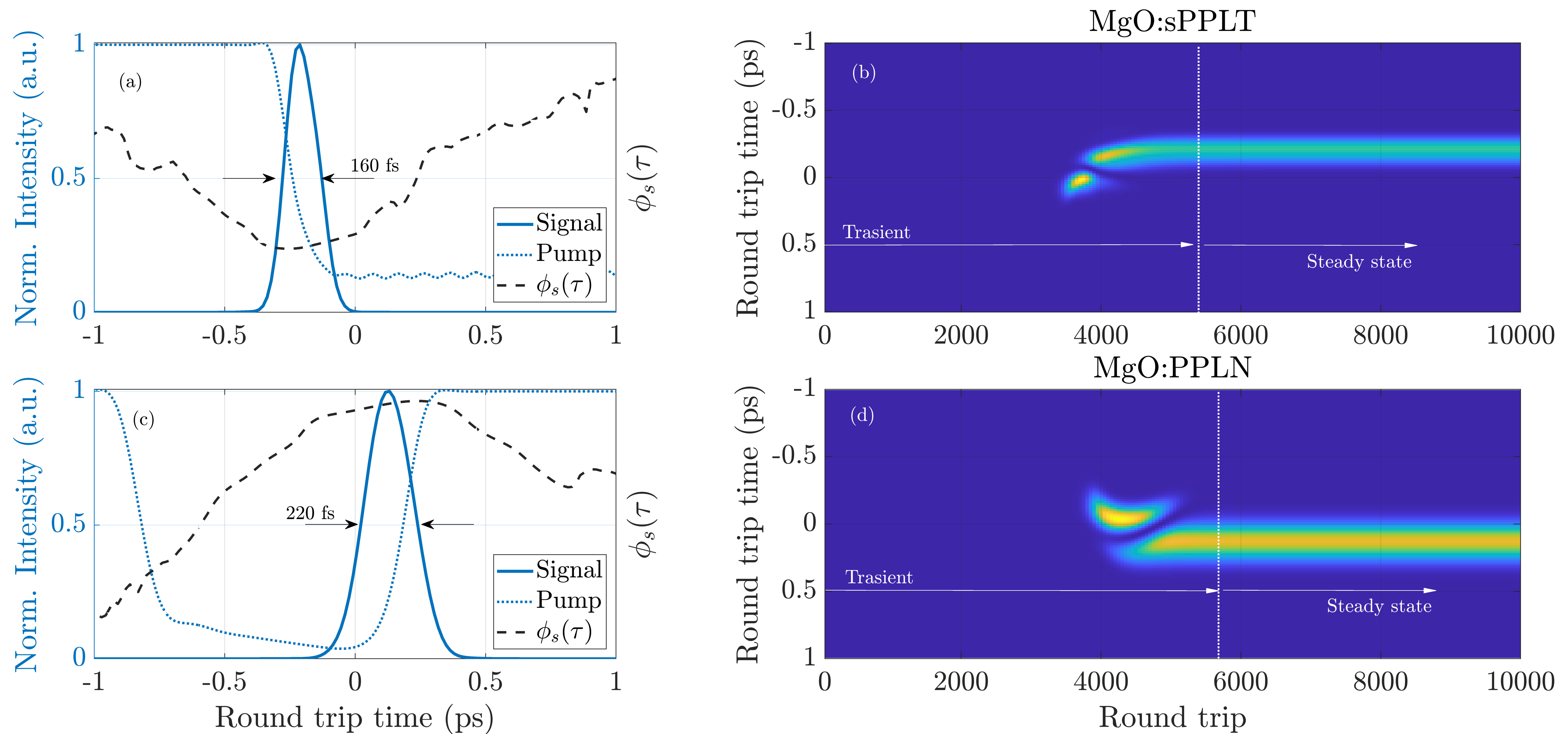}
    \caption{Intracavity quadratic soliton formation for $\delta f= \delta f^{\mathrm{LB}}$. Panels (a,c) show the signal and pump normalized intensities and the signal phase for a short-time window of the full round trip, for both normal and anomalous dispersion regimes. Panels (b,c) show the signal intensity for the full simulation, where both transient and steady states are indicated.}
    \label{fig:time1}
\end{figure*}

As mentioned in Section~\ref{sec:freqdomain}, a proper fine-tuning of phase modulation frequency results in a single structure due to the threshold condition tuning. Figure~\ref{fig:time1} shows the numerical solutions for the temporal domain by setting the appropriate $\delta f = \delta f^{\mathrm{LB}}$ to achieve the single intracavity pulse corresponding to a solitonic solution. Panels (a, c) show the intracavity quadratic soliton along with single-pass pump intensity and the signal phase for both nonlinear crystals, exhibiting pulse duration of 160 and 220~fs, respectively. Panels (b, d) show the evolution of the signal intensity along 10$^3$ round-trips. As can be seen, after the $\approx 5000$-th round-trip the signal intensity reaches the steady state. This is the result of the balance between the parametric gain provided by the nonlinear crystal, the intracavity EOM with the proper frequency adjustment, and the GDD compensation. Similar to the case of two pulses, the change in the temporal phase along the signal pulse in this configuration is $\Delta\phi_s \approx 0.3$~rad, showing again that our system is capable of generating nearly transform-limited pulses, with a coherent and broadband spectrum corresponding to a frequency comb in both normal and anomalous dispersion regimes.

%%%%%%%%%%%%%%%%%%%%%%%%%%%% CONCLUSIONS %%%%%%%%%%%%%%%%%%%%%%%%%%%%%%%%%
\section{Conclusions}
In conclusion, we have presented a new class of frequency comb based on phase-modulated bulk cw-driven OPO with dispersion control. We have theoretically shown that our proposed scheme provides a coherent phase-locked spectrum that serves as a frequency comb and admits as a solution single intracavity quadratic soliton in the time domain, in both normal and anomalous dispersion regimes. We have presented the conditions for optimum frequency detuning to generate quadratic solitons, by tailoring the threshold condition in the temporal domain. Our results pave the way for the realization of a new class of frequency comb with excellent temporal and spectral characteristics exploiting the vast potential of OPOs using widely available cw pump lasers.

%%%%%%%%%%%%%%%%%%%%%%%%%%%% CONCLUSIONS %%%%%%%%%%%%%%%%%%%%%%%%%%%%%%%%%
\begin{acknowledgments}
We gratefully acknowledge funding from the Ministerio de Ciencia e Innovación (MCIN) and the State Research Agency (AEI), Spain (Project Nutech PID2020-112700RB-I00); Project Ultrawave EUR2022-134051 funded by MCIN/AEI and by the “European Union NextGenerationEU/PRTR; Severo Ochoa Programme for Centres of Excellence in R\&D (CEX2019-000910-S);  Generalitat de Catalunya (CERCA); Fundación Cellex; Fundació Mir-Puig. S. Chaitanya Kumar acknowledges support of the Department of Atomic Energy, Government of India, under Project Identification No. RTI 4007. 
\end{acknowledgments}

\nocite{*}

\bibliography{apssamp}

\end{document}